\documentclass[aps,prl,twocolumn,showpacs,preprintnumbers,amsmath,amssymb]{revtex4-1}
\usepackage{graphicx}
\usepackage{color}
\usepackage[normalem]{ulem}  

\renewcommand\sout{\bgroup \color{red}\ULdepth=-.5ex \ULset}
\newcommand\soutb{\bgroup \color{blue} \ULdepth=-.5ex \ULset}

\usepackage{graphicx}
\usepackage{dcolumn}
\usepackage{bm}

\begin{document}

\preprint{}

\title{Unobservability of topological charge in nonabelian gauge theory}

\author{Nodoka~Yamanaka$^{1,2}$}
  \affiliation{$^1$Kobayashi-Maskawa Institute for the Origin of Particles and the Universe, Nagoya University, Furocho, Chikusa, Aichi 464-8602, Japan}
  \affiliation{$^2$Nishina Center for Accelerator-Based Science, RIKEN, Wako 351-0198, Japan}
  \email{nyamanaka@kmi.nagoya-u.ac.jp}

\date{\today}

\begin{abstract}
We show that the topological charge of nonabelian gauge theory is unphysical by using the fact that it always involves the unphysical gauge field component proportional to the gradient of the gauge function.
The removal of Gribov copies, which may break the Becchi-Rouet-Stora-Tyutin symmetry, is irrelevant thanks to the perturbative one-loop finiteness of the chiral anomaly.
The unobservability of the topological charge immediately leads to the resolution of the Strong CP problem.
We also present important consequences such as the physical relevance of axial $U(1)$ symmetry, the $\theta$-independence of vacuum energy, the unphysicalness of topological instantons, and the impossibilities of realizing the sphaleron induced baryogenesis as well as the chiral magnetic effect.
The unphysical vacuum angle and the axial $U(1)$ symmetry also imply that the CP phase of the Cabibbo-Kobayashi-Maskawa matrix is the sole source of CP violation of the standard model.
\end{abstract}


\maketitle

The nonabelian gauge theory is known to have nontrivial topology, and it has been very widely investigated in the past.
It is known that the four-dimensional space-time integral of the topological charge density
\begin{equation}
\frac{\alpha_s}{8\pi}
F_{\mu \nu, a}\tilde F^{\mu \nu}_a
,
\label{eq:topologicalchargedensity}
\end{equation}
may not vanish, in spite of being a total divergence.
Here $F$ and $\tilde F$ are the strengths of nonabelian gauge field and its dual, with the coupling constant $\alpha_s$.
Its study has been initiated by the discovery of the instanton solution in gauge theories \cite{Belavin:1975fg,Callan:1977gz,Schafer:1996wv}, and the connection with the chiral anomaly \cite{Adler:1969gk,Bell:1969ts,Fujikawa:1979ay,Fujikawa:1980eg} lead to claim that the absence of the $U(1)_A$ Nambu-Goldstone (NG) boson (the so-called $U(1)_A$ problem) is resolved \cite{tHooft:1976rip,tHooft:1976snw}, although several other possibilities were also proposed \cite{Witten:1979vv,Veneziano:1979ec,Veneziano:1980xs,Kogut:1973ab,Kogut:1974kt,Kugo:1978nc}.
In recent lattice simulations, results consistent with the restoration of the $U(1)_A$ symmetry at finite temperature in QCD were reported \cite{Cossu:2013uua,Brandt:2016daq,Aoki:2020noz}, 
and the explicit breaking of $U(1)_A$ is still a matter of debate. 
On the other hand, the existence of distinct topological sectors in nonabelian gauge theory posed another serious problem.
To avoid the violation of the cluster decomposition principle by operators carrying topological or axial charges, the coherent superposition of all topological sectors with a vacuum angle $\theta$ was claimed to be required \cite{Jackiw:1976pf,Callan:1976je}.
However, a theory with such vacuum has a term proportional to Eq. (\ref{eq:topologicalchargedensity}) in the Lagrangian, the so-called $\theta$-term, and it is in general CP violating.
The null experimental results of electric dipole moment (EDM) are currently pointing out to us an extreme fine-tuning, $|\theta|<10^{-10}$, which is also named as the Strong CP problem \cite{Crewther:1979pi,Yamanaka:2017mef,Chupp:2017rkp,Abel:2020gbr,Graner:2016ses}.
Several resolutions have been proposed in the past, such as the axion mechanism \cite{Peccei:1977hh}, the spontaneous breaking of exact CP symmetry of ultraviolet physics \cite{Nelson:1983zb,Barr:1984qx,Barr:1984fh}, or adjusting one of the quark mass to zero, which is now in conflict with lattice calculations \cite{Alexandrou:2020bkd}.

In this work, we propose a resolution to these problems by finding an intrinsic mechanism which makes unphysical the topological charge.
Our findings also have several important byproducts in particle physics and cosmology, such as the physical irrelevance of the chiral magnetic effect, the baryon/lepton number violation in electroweak gauge theory \cite{tHooft:1976snw}, and the restoration of $U(1)_A$ symmetry at high temperature.

The key observation is that the topological charge always involves the longitudinal component of the gauge field.
The local gauge symmetry 
\begin{equation}
A_a^\mu (x) \to A_a^\mu (x) + \partial^\mu \chi_a (x) + O(g_s)
,
\label{eq:localgaugetransform}
\end{equation}
implies that the longitudinal polarization of the gauge field $A$ is unphysical in the perturbative expansion of quantum field theory as a consequence of the Becchi-Rouet-Stora-Tyutin (BRST) symmetry \cite{Becchi:1975nq,Tyutin:1975qk}, either by vanishing intrinsically, or cancelling with the Faddeev-Popov ghost at the level of observables \cite{Gupta:1949rh,Bleuler:1950cy,Kugo:1977zq,Kugo:1979gm}.
Remaining terms that are higher order in the gauge coupling $g_s = \sqrt{4 \pi \alpha_s}$ may upset the above statement, but this only happens beyond the perturbation theory.

Let us show that the topological charge contains the longitudinal polarization.
It is well known that the integral of Eq. (\ref{eq:topologicalchargedensity}) yields an integer number:
\begin{eqnarray}
&&
\frac{\alpha_s}{8\pi}
\int d^4 x\,
F_{\mu \nu, a}\tilde F^{\mu \nu}_a
\nonumber\\
&=&
\frac{i g_s \alpha_s}{24\pi} 
\int d^3 \vec{x}\, 
f_{abc} \epsilon_{ijk} A_{ia} (\vec{x}) A_{jb} (\vec{x}) A_{kc} (\vec{x})
\bigg|_{t=-\infty}^{t=+\infty} 
\nonumber\\
&=&
\Delta n
,
\label{eq:topologicalcharge}
\end{eqnarray}
where $\Delta n$ is the change of the winding number which labels the topological sector.
We see, however, that the first equality of the above equation has a triple product (contraction with $\epsilon_{ijk}$) of the spatial components of the gauge field.
Since $\epsilon_{ijk}$ must be contracted with three vectors with different spatial directions each other, the triple product is proportional to all components of the gauge field, especially to the 3-dimensional gradient $\vec \nabla \chi_a (\vec x)$, at any given spatial point. 
We note that in the first equality of Eq. (\ref{eq:topologicalcharge}) the temporal integral has been performed so that the remaining expression has no time-dependence (at $t=\pm \infty$).
There the four-derivative of the gauge function becomes $\vec \nabla \chi_a (\vec x)$, and the contraction of three gauge field operators with $\epsilon_{ijk}$ is always proportional to this gradient, which corresponds to the longitudinal polarization or momentum after Fourier transform.
After Lorentz covariantly fixing the gauge (e.g. $\partial_\mu A_a^\mu =0$), the gauge function $\chi_a$ turns to a Faddeev-Popov ghost field.
According to the local gauge/BRST symmetries, any processes containing longitudinally polarized gauge fields and/or ghosts are unphysical \cite{Gupta:1949rh,Bleuler:1950cy,Kugo:1977zq,Kugo:1979gm}, then so do those involving the topological charge, even if they are finite at the amplitude level.
Since the topological charge operator has no BRST daughter operator, it must vanish at the observable level only by summing over states, namely over all topological sectors
\begin{eqnarray}
\sum_{| \Omega \rangle \ne | 0 \rangle}
\xi \langle 0| F_{\mu \nu ,a} \tilde F^{\mu \nu}_a (x) |\Omega \rangle \langle \Omega | F_{\rho \sigma ,b} \tilde F^{\rho \sigma}_b (x) | 0\rangle
&=&
0
,
\end{eqnarray}
where $\xi$ is an arbitrary constant.
We make an important emphasis that the topology changing contribution of observables (amplitude squared) may always be factorized in this form, since if the topological charge density operator starts to have correlations with other operators, their operator products will never generate the topological charge itself, according to Adler-Bardeen theorem which forbids contributions from radiative corrections to Eq. (\ref{eq:topologicalchargedensity}) \cite{Adler:1969er,Adler:1972zd,Higashijima:1981km,Costa:1977pd,Fujikawa:1981rw,Lucchesi:1986sp,Anselmi:2014kja,Anselmi:2015sqa,Mastropietro:2020bhz}.
This perturbative one-loop finiteness is a nonperturbative property which is valid at all scale, and it is part of the anomaly matching condition of 't Hooft \cite{thooftanomalymatchingconodition}.
This then implies that the change of topological sectors does not contribute to observables.
A more robust derivation based on the Ward-Takahashi identity (WTI) will be given in a full paper \cite{fullpaper}.

So far we assumed the BRST symmetry and the unphysicalness of the longitudinal component of the gauge field which both only hold in perturbation theory.
The BRST symmetry breaks down when the path integral is restricted to the first Gribov region for removing Gribov copies \cite{Gribov:1977wm,Singer:1978dk,Zwanziger:1989mf,Maggiore:1993wq,Dudal:2008sp,Vandersickel:2012tz} (and most probably to the fundamental modular region which removes all Gribov copies, but for which the derivation is still not established),
 and this might upset our arguments.
The Gribov copies are gauge configurations which fulfill the same gauge fixing condition while being connected to each other via continuous gauge transformations, and they occur due to the nonabelian gauge self-interactions at sufficiently large gauge field amplitude (not to be confused with the amplitude of quantum processes).
The restriction of the path integral applies to the amplitude of the fields so as to ideally only contain one gauge configuration per gauge orbit.
It is actually known that the gauge bosons get mass in this circumstance \cite{Dudal:2008sp,Vandersickel:2012tz}, which is also seen in explicit lattice calculations with Landau gauge fixing \cite{Iritani:2009mp,Dudal:2018cli,Falcao:2020vyr}.
We will from now show that for the present case, the BRST symmetry and the unobservability of the topological charge still hold.
The chiral anomaly, also given by Eq. (\ref{eq:topologicalchargedensity}), is actually known to not be renormalized, and it is strictly perturbation finite, again according to Adler-Bardeen theorem \cite{Adler:1969er,Adler:1972zd,Higashijima:1981km,Costa:1977pd,Fujikawa:1981rw,Lucchesi:1986sp,Anselmi:2014kja,Anselmi:2015sqa,Mastropietro:2020bhz}.
It then always appears perturbatively in quantum processes, even if the gauge fields may be corrected nonperturbatively.
An important property of perturbatively defined field operators (i.e. we may count the number of field quanta) is that fields are generated with infinitesimal amplitude according to Glauber's coherent field formulation \cite{Glauber:1963tx},
\begin{equation}
| \epsilon \rangle
=
e^{-\frac{1}{2}|\epsilon|^2}
\sum_n
\frac{(\epsilon a^\dagger)^n}{n!}
| 0 \rangle
=
(1+ \epsilon a^\dagger)| 0 \rangle
,
\end{equation}
where $\epsilon$ is the infinitesimal amplitude of a field created by the operator $a^\dagger$ (the term with $1$ has no physical effect since it does not create any fields).
Infinitesimal field configurations always lie in the fundamental modular region and the BRST symmetry is exact in perturbation, so the chiral anomaly and the topological charge, both given by Eq. (\ref{eq:topologicalchargedensity}), are not affected by the Gribov ambiguity.
We may therefore safely claim that the topological charge is unphysical.

The most important outcome of the above discussion is that the $\theta$-term, which expresses the coherent superposition of all topological sectors, is unobservable.
This is also true for the classically CP conserving $\theta = \pi$ case, where the spontaneous breaking of CP symmetry occurs (Dashen's phenomenon) \cite{Dashen:1970et,Gaiotto:2017yup}.
Let us note that we have no more concern with the cluster decomposition since the topology change, although it does not damp at large spatial separation between operators, is now unphysical.
This then implies that {\it the Strong CP problem is resolved}.
Since the $\theta$-term and topological sectors are unphysical, the vacuum energy must be $\theta$-independent.
The vacuum energy is in principle an $O(1)$ quantity in the $1/N_c$ expansion, and it presented so far a paradox when massless quarks are introduced, since it was believed that the $\theta$-dependence is rotated away by $O(1/N_c)$ quark contributions \cite{Witten:1979vv,Veneziano:1979ec,Veneziano:1980xs}.
Our conclusion resolves this problem in a trivial way, since the vacuum energy has no $\theta$-dependence.
We may also infer that the topological instantons, generated by classical self-dual configurations $\tilde F_{\mu \nu} =\pm F_{\mu \nu}$, become unphysical.
Another nontrivial consequence is that the $U(1)_A$ symmetry also becomes relevant in the physical subspace, since the explicit breaking of $U(1)_A$ by the topological charge is now unphysical.
We will next discuss the latter two topics in detail by introducing fermions.
We also note that, our statement will in principle allow one to choose a distinct topological sector labeled by a definite winding number in lattice simulations, and may alleviate the computational cost for cases where fixed topology is computationally advantageous.

We now introduce fermions and inspect in detail the chiral (or axial) WTI \cite{Bardeen:1969md}
\begin{equation}
\sum_{\psi}^{N_f}
\Bigl[
\partial^\mu (\bar \psi \gamma_\mu \gamma_5 \psi )
+2m_\psi
\bar \psi i\gamma_5 \psi
\Bigr]
=
-
\frac{N_f\alpha_s}{8\pi}
F_{\mu \nu, a}\tilde F^{\mu \nu}_a
,
\label{eq:chiralWTI}
\end{equation}
where the fermions $\psi$ are summed over all active flavors $N_f$.
We previously saw that the topological charge is unobservable.
There should then also be an unphysical contribution on the left-hand side of Eq. (\ref{eq:chiralWTI}).
From Atiyah-Singer's index theorem which relates the zero-modes of the Dirac operator $D\hspace{-0.65em}/\, \equiv \partial \hspace{-0.5em}/\, -ig_s A \hspace{-0.55em}/\,_a t_a$ to the topological charge
\begin{equation}
{\rm ind}(D\hspace{-0.65em}/\,)
=
-\frac{N_f \alpha_s}{8\pi}
\int d^4 x\,
F_{\mu \nu, a}\tilde F^{\mu \nu}_a
,
\label{eq:atiyah-singer}
\end{equation}
where ${\rm ind}(D\hspace{-0.65em}/\,)$ is the difference between the numbers of Dirac zero-modes with positive (right-handed) and negative (left-handed) chiralities, we infer that the contribution from chiral Dirac zero-modes is the unphysical piece.
By removing these unobservable parts from Eq. (\ref{eq:chiralWTI}), we end up with the ``physical'' chiral WTI:
\begin{eqnarray}
\sum_{\psi}^{N_f}
\Bigr[
\partial^\mu (\bar \psi \gamma_\mu \gamma_5 \psi )
+2m_\psi
\bar \psi i\gamma_5 \psi
\Bigr]_{\lambda \ne 0}
\nonumber\\
=
-\frac{N_f\alpha_s}{8\pi} F_{\mu \nu, a}\tilde F^{\mu \nu}_a
\Bigr|_{\Delta n =0}
,
\label{eq:physicalchiralWTI}
\end{eqnarray}
where the subscript of the left-hand side $\lambda \ne 0$ means that we removed the chiral Dirac zero-modes.
We note that Eq. (\ref{eq:physicalchiralWTI}) still locally violates the $U(1)_A$ symmetry due to the $F\tilde F$ term, but this violation disappears in the low energy-momentum limit, since it is a total divergence.
This means that QCD and other vectorlike nonabelian gauge theories with fermions are $U(1)_A$ symmetric up to the current fermion mass $m_\psi$, and they will suffer from the spontaneous chiral symmetry breaking, just like the other flavor nonsinglet axial symmetries.

Let us check whether the dynamical $U(1)_A$ symmetry breaking of massless QCD truly realizes the well-known symmetry of hadron physics or not.
The global symmetry of the quark sector is \cite{Fukushima:2010bq,Tanizaki:2018wtg}
\begin{eqnarray}
G^{\rm (QCD)}
&=&
U(N_f)_L \times U(N_f)_R / \mathbb{Z}_{N_c}
\nonumber\\
&=&
\frac{SU(N_f)_L \times SU(N_f)_R \times U(1)_V \times U(1)_A}{\mathbb{Z}_{N_c} \times (\mathbb{Z}_{N_f})_L \times (\mathbb{Z}_{N_f})_R \times \mathbb{Z}_2}
, \ \ \ \ 
\end{eqnarray}
where we used $U(N_f) = \frac{SU(N_f) \times U(1)}{\mathbb Z_{N_f}}$ and $U(1)_L \times U(1)_R = \frac{U(1)_V \times U(1)_A}{\mathbb Z_2}$.
The chiral symmetry breaking pattern of QCD is
\begin{eqnarray}
\frac{SU(N_f)_L \times SU(N_f)_R}{(\mathbb{Z}_{N_f})_L \times (\mathbb{Z}_{N_f})_R} 
&\to & 
\frac{SU(N_f)_V }{(\mathbb{Z}_{N_f})_V }
,
\\
U(1)_A
&\to & 
\mathbb{Z}_2 
,
\end{eqnarray}
where this time $U(1)_A$ is also spontaneously broken (the resulting $\mathbb{Z}_2 $ is because $U(1)_A$ contains the $U(1)_V$ element $e^{i\gamma_5 \pi } = e^{i\pi } = -1$).
We finally obtain the well-known symmetry of hadron physics $G^{\rm ({\rm hadron})}=\frac{SU(N_f)_V \times U(1)_B }{(\mathbb{Z}_{N_f})_V}$, where $U(1)_B = U(1)_V / \mathbb{Z}_{N_c}$ is due to the baryon which contains $N_c$ confined quarks \cite{Tanizaki:2018wtg}.

The relevance of the $U(1)_A$ symmetry suggests us that it will be restored at high temperature.
This statement is actually totally consistent with recent lattice results \cite{Cossu:2013uua,Brandt:2016daq,Aoki:2020noz}.
This also implies that $\eta'$ and $\eta$, despite the contribution from the chiral anomaly (\ref{eq:physicalchiralWTI}), are (pseudo) NG bosons.
We then predict that $\eta'$ becomes massless as the up, down, and strange quark masses tend to zero.
This may be tested with lattice QCD in the future.

Eq. (\ref{eq:physicalchiralWTI}) also implies that the $U(1)_A$ complex phase rotation of fermion masses does not anymore affect the $\theta$-term and vice versa, as the $\theta$-term is unphysical.
The CP phases of the quark mass then become part of the field definition, so the CP violation of the standard model (SM) is only given by the complex phase of the Cabibbo-Kobayashi-Maskawa (CKM) matrix \cite{Kobayashi:1973fv}.
At the same time, constraints on new physics beyond the SM given by EDM experiments are significantly relaxed, without introducing additional fields such as the axions.

Let us now see the implication of the unobservability of chiral Dirac zero-modes.
It was so far believed that the $U(1)_A$ symmetry, explicitly broken by the anomaly, generates the 't Hooft vertex, an effective contact multi-fermion interaction \cite{tHooft:1976rip,tHooft:1976snw}.
For instance, in 3-flavor QCD with broken $U(1)_A$, we have the following interaction at low energy (the so-called Kobayashi-Maskawa-'t Hooft interaction \cite{Kobayashi:1970ji,Kobayashi:1971qz,Hatsuda:1994pi})
\begin{equation}
{\cal L}_{\rm KMT}
=
C \bar u_R u_L \bar d_R d_L \bar s_R s_L + {\rm h.c.}
,
\label{eq:KMT}
\end{equation}
where $C$ is a phenomenological parameter which may be obtained from the analysis of instantons.
This three-quark force is $SU(3)_L \times SU(3)_R$ invariant, and it is generated by the chiral Dirac zero-modes \cite{tHooft:1976rip,tHooft:1976snw}.
Since the latters are unphysical, this multi-quark interaction should be as well.
This is consistent with the fact that no direct and obvious effects of the 't Hooft vertex have been observed in collider experiments \cite{Khoze:2019jta,Khoze:2020tpp,Khoze:2020paj} as well as with the null experimental result in the search for the notorious chiral magnetic effect \cite{Kharzeev:2007jp,Fukushima:2008xe,Kharzeev:2015znc,STAR:2021pwb} which was expected to be an excellent probe of the chiral chemical potential generated by the topological charge.
The 't Hooft vertex is also known to help the stability of the neutron star by making a smooth crossover regarding the baryon density in QCD at low temperature \cite{Hatsuda:2006ps,Abuki:2010jq,Baym:2017whm}, so its removal will certainly affect the neutron star physics, and careful studies will definitely be needed in the future.

The most important zero-mode induced multi-fermion interaction to be discussed is the baryon ($B$) and lepton ($L$) number violating one generated by the $U(1)_{B+L}$ anomaly of the electroweak theory.
Processes like $u + d \to \bar d + \bar s + 2\bar c +3\bar t + e^+ + \mu^+ + \tau^+$ were actually expected to happen via the sphaleron process \cite{Manton:1983nd,Klinkhamer:1984di,Fukugita:1986hr} at high temperature in the early universe, and to eventually explain the baryogenesis \cite{Sakharov:1967dj,Riotto:1999yt}.
According to our discussion, this anomalous $B+L$ violation is physically forbidden.
We therefore have to think of particle physics scenarios with explicit baryon number violating interactions \cite{Georgi:1974sy,Affleck:1984fy,Barbier:2004ez,Dorsner:2016wpm} to explain the matter abundance around us.

To summarize, we showed that the topological charge and the chiral Dirac zero-modes are unphysical in nonabelian gauge theory.
Let us list the important consequences of this finding:
\begin{itemize}
\item
Resolution of the Strong CP problem.

\item
The vacuum energy does not depend on $\theta$.

\item
Topological instantons are unphysical.

\item
The $U(1)_A$ symmetry is not broken by the anomaly.

\item
The CP violation of the SM is only given by the CP phase of the CKM matrix.

\item
The chiral magnetic effect does not occur.

\item
The sphaleron induced baryogenesis does not occur.

\end{itemize}
There are several other interesting topics that could not be covered or discussed in detail in this presentation, especially the phenomenological consequences.
We leave their inspection to another paper \cite{fullpaper}.
We also saw that the consistency of our arguments with lattice QCD, regarding e.g. the simulation with fixed topology or the chiral extrapolation of $m_{\eta'}$, will have to be checked.

\begin{acknowledgments}
The author thanks Yoshikazu Hagiwara, David Dudal, and Hiroaki Abuki for useful discussions.
This work was supported by Daiko Foundation.
\end{acknowledgments}

\end{document}